\documentstyle[preprint,revtex]{aps}
\begin{document}
\draft
\preprint{Alberta Thy--26--93}
\begin{title}
A Bjorken sum rule for semileptonic $\Omega_b$ decays to
ground and excited charmed baryon states
\end{title}
\author{Q. P. Xu }
\begin{instit}
Theoretical Physics Institute and Physics Department\\
University of Alberta\\
Edmonton, Alberta, Canada, T6G 2J1
\end{instit}
\begin{abstract}
We derive a Bjorken sum rule for semileptonic $\Omega_b$ decays to
ground and low-lying negative-parity excited charmed baryon states,
in the heavy quark limit.
We discuss the restriction from this sum rule on
form factors and compare it with some models.
\end{abstract}
\narrowtext

In the heavy quark limit, $m_{b,c}>>\Lambda_{QCD}$,
a new $SU(2N)$ ($N=$number of heavy
flavors) spin-flavor symmetry appears in QCD \cite{r1}.
Consequently, all the form factors involved in
$B \rightarrow (D, D^\ast) e {\bar\nu}_{e}$,
can be expressed by only one universal form factor,
the Isgur-Wise function, which is normalized at the maximum
momentum transfer, $q^2_m$ \cite{r1}. This is also true for
$\Lambda_b \rightarrow \Lambda_c l \bar \nu_{l}$ \cite{r1}.
For $\Omega_b \rightarrow (\Omega_c, \Omega_c^\ast) e \bar \nu_e$,
the situation is a little more complicated.
In the heavy quark limit there are two independent
form factors \cite{r2,r3},
since the light components (light quarks and gluons)
in the baryons have spin 1. One of
these two form factors is normalized at $q^2_m$. It is known
that in the heavy quark limit,

the Lorentz structure of the amplitude for
$\Sigma_b \rightarrow (\Sigma_c, \Sigma_c^\ast) e \bar \nu_e$
are exactly the same as those for
$\Omega_b \rightarrow (\Omega_c, \Omega_c^\ast) e \bar \nu_e$
\cite{r2}.

In the heavy quark limit, one can also derive the so-called
Bjorken sum rule for semileptonic decays of bottom hadrons
to charmed states \cite{r5,r6}. Such a sum rule has already been
given for $B$ and $\Lambda_b$ semileptonic decays \cite{r5,r6}.
Taking $B$-meson semileptonic decays as an example, the sum rule
follows from the consideration of the quantity \cite{r6}
\FL
\begin{eqnarray}
&&\!\!h^{X_c}_\Gamma (w)\!\!=\!\!\sum_{s,s^\prime}
\langle B (v,s) | \bar b \Gamma c | X_c (v^\prime, s^\prime)
\rangle \langle X_c (v^\prime, s^\prime) | \bar c \Gamma b |
B (v,s) \rangle\nonumber\\
&& \
\label{eq:e1}
\end{eqnarray}
where $s$ and $s^\prime$ indicate the spin indices, $w$
is the velocity product ($w=v\cdot v^\prime$) and is limited to be
of order unity. If one sums over the final charmed hadronic states
$X_c$ with masses from $m_D$ to $m_D+\mu$, then
if $\mu >>\Lambda_{QCD}$, this inclusive quantity can be computed
in the perturbative QCD as a heavy quark transition. On the other
hand, if $\mu <<m_D$, the quantity $h^X_\Gamma (w)$ can be
calculated for exclusive decays using the heavy quark effective
theory \cite{r1}. Thus \cite{r6}
\FL
\begin{eqnarray}
\frac{|C_{ji}(\mu)|^2}{2}
Tr \lbrace (\not\!v^\prime+1)\bar \Gamma (\not\!v+1) \Gamma
\rbrace
=\!\!\sum_{ m_{X_c}-m_D<\mu } h^{X_c}_\Gamma (w)
\label{eq:e2}
\end{eqnarray}
where $\bar \Gamma=\gamma_0 \Gamma^+ \gamma_0$.
The factor $C_{ji}(\mu)$ arises from the perturbative QCD
correction to the heavy quark transition summed in the leading
logarithmic approximation. One obtains \cite{r6}
\FL
\begin{eqnarray}
1=\frac{(w+1)}{2}|\xi(w)|^2+...
\label{eq:e3}
\end{eqnarray}
where $\xi(w)$ is the Isgur-Wise function for
$B \rightarrow (D, D^\ast)$ transitions and
the ellipsis denotes the contributions from excited states.
For brevity we have not shown the complete form
of the sum rule given in \cite{r6}. In Eq.~(\ref{eq:e3})
a factor of $|C_{ji}(\mu)|^2$ from both sides of
Eq.~(\ref{eq:e2}) has been cancelled so that the form factors such
as $\xi(w)$ can be evaluated at $\mu=\mu_c$.
The sum rule in Eq.~(\ref{eq:e3}) also supplies an upper limit to
$\xi(w)$ since the contributions from the excited
states in Eq.~(\ref{eq:e3}) is always positive.

In this note we will follow a method parallel
to \cite{r6} to derive a Bjorken
sum rule for $\Omega_b$ semileptonic decays to ground and
low-lying negative-parity excited charmed baryon states.
We will also discuss the upper limit on form factors coming
from this sum rule and compare it with existing models.

To begin with, we recall the Lorentz structure of the amplitude for
$\Omega_b$ semileptonic decays to ground states, $\Omega_c$
and $\Omega_c^\ast$, in the heavy quark limit, using the tensor
method \cite{r3,r4}. Since the spin of light component in these
baryons is also conserved in the heavy quark limit, one needs to
build a tensor which carries the indices of both the heavy quark
spin and the spin of the light components. Thus \cite{r3}
\FL
\begin{eqnarray}
\langle \Omega_c\  {\rm or}\ \Omega_c^\ast&&|
\bar b\ \Gamma\ c | \Omega_b \rangle\!\!=\!\!
{\bar B}^\mu_c (v^\prime)\ \Gamma B^\nu_b(v)
(\!-\xi_1\ g_{\mu\nu}+\xi_2\ v^\mu v^{\prime\nu} )\nonumber\\
&& \
\label{eq:e4}
\end{eqnarray}
with
\FL
\begin{eqnarray}
&&B^\mu_c(v^\prime)=
\frac{(\gamma^\mu+v^{\prime\mu}) \gamma_5 }{\sqrt3}
u_{\Omega_c}(v^\prime)\ \ (for\ \Omega_c)\nonumber\\
&&B^\nu_b(v)=\frac{(\gamma^\nu+v^{\nu}) \gamma_5 }{\sqrt3}
u_{\Omega_b}(v) \ \ (for\ \Omega_b)\nonumber\\
&&B^\mu_c(v^\prime)=u^\mu_{\Omega_c^\ast}(v^\prime)
\;\;\;\;\;\;\;\;\;\;\;\;\;\;\;\;\;\;\;\;(for\ \Omega_c^\ast)
\label{eq:e5}
\end{eqnarray}
where $u$ is a spinor and $u^\mu$ is a Rarita-Schwinger spinor
vector for a spin-3/2 particle.
The indices indicate the corresponding baryons.
The form factors $\xi_1$ and $\xi_2$ are functions of
$w$. $\xi_1$ is normalized at $w=1$ \cite{r2,r3}:
\begin{eqnarray}
\xi_1(1)=1
\label{eq:e6}
\end{eqnarray}
Note the relations between $\xi_{1,2}$ and the form factors
$\eta$ and $ \tau$ used in \cite{r2} are:
$\xi_1=\eta-\frac{(w-1)}{2}\iota$, $\xi_2=-\frac{1}{2}\iota$.

To study decays to excited charmed states,
we first consider the qualitative
characteristics of the low-lying negative parity excited states.
Since the heavy quark symmetry cannot tell us anything about the
spectroscopy of the light degrees of freedom, we will use
constituent quark model as a guide to classify the excited states.
We consider charmed baryon states formed by $(css)$.
We use $s_{ss}$, $l_r$ and $l_{ss}$ to denote the spin,
the relative angular momentum and the center-of-mass angular
momentum of the two $s$-quarks, respectively. Then the ground
states have $(s_{ss}, l_r, l_{ss})=(1, 0, 0)$. If we write
$s_l=s_{ss}+l_r+l_{ss}$, the light quarks in ground states
have total angular momentum
and parity $s_l^{\pi_l}=1^+$. The corresponding ground state
baryons are $\Omega_c$ and $\Omega_c^\ast$. As for the lowest
negative-parity
excited states, it could either have the angular momentum
excitation in $l_{ss}$ and thus $(s_{ss}, l_r, l_{ss})=(1, 0, 1)$ and
$s_l^{\pi_l}=2^-, 1^-, 0^-$,
or have the angular excitation in $l_r$ and thus
$(s_{ss}, l_r, l_{ss})=(0, 1, 0)$ and $s_l^{\pi_l}=1^-$. On the
baryon level, we denote these excited states with
${\tilde \Omega}_c (S^\pi, s_l^{\pi_l})$ where
$S^\pi$ is the total spin and parity of the baryon. Thus
corresponding to $(s_{ss}, l_r, l_{ss})=(1, 0, 1)$, there are
${\tilde \Omega}_c (5/2^-, 2^-), {\tilde \Omega}_c (3/2^-, 2^-),
{\tilde \Omega}_c (3/2^-, 1^-), {\tilde \Omega}_c (1/2^-, 1^-)$
and ${\tilde \Omega}_c (1/2^-, 0^-)$. Corresponding to
$(s_{ss}, l_r, l_{ss})=(0, 1, 0)$, there are
${\tilde \Omega}_c (3/2^-, 1^-)$ and
${\tilde \Omega}_c (1/2^-, 1^-)$.
We will, however, not consider the negative-parity excited states
corresponding to $(s_{ss}, l_r, l_{ss})=(0, 1, 0)$ in the
discussion to follow. The reason is as follows:
In a spectator quark model the light quarks are spectators in
transitions of $\Omega_b$ to these excited charmed states.
Since the light
quarks in $\Omega_b$ and in these excited states have different
intrinsic spins, the corresponding form factors vanish. Thus it is
quite safe to expect the contributions from these excited states
would be very small. Also it is expected in a non-relativistic quark
model \cite{r6b}
that the masses of the excited states corresponding to
$(s_{ss}, l_r, l_{ss})=(0, 1, 0)$ are higher than those
corresponding to $(s_{ss}, l_r, l_{ss})=(1, 0, 1)$.
Therefore we will confine our discussion
only to excited states corresponding to
$(s_{ss}, l_r, l_{ss})=(1, 0, 1)$, i.e.
${\tilde \Omega}_c (5/2^-, 2^-)$, ${\tilde \Omega}_c (3/2^-, 2^-)$,
${\tilde \Omega}_c (3/2^-, 1^-)$, ${\tilde \Omega}_c (1/2^-, 1^-)$
and ${\tilde \Omega}_c (1/2^-, 0^-)$.

The form factor structures in semileptonic $\Omega_b$ decays to
${\tilde \Omega}_c (5/2^-, 2^-)$, ${\tilde \Omega}_c (3/2^-, 2^-)$,
${\tilde \Omega}_c (3/2^-, 1^-)$, ${\tilde \Omega}_c (1/2^-, 1^-)$
and ${\tilde \Omega}_c (1/2^-, 0^-)$ are quite
simple in the heavy quark limit. It is rather easy to
count the number of form factors in these decays using
the helicity analysis \cite{r7}. One finds that for
$\Omega_b$ decays to ${\tilde \Omega}_c (5/2^-, 2^-)$ and
${\tilde \Omega}_c (3/2^-, 2^-)$, there are two form factors.
For decays to ${\tilde \Omega}_c (3/2^-, 1^-)$ and
${\tilde \Omega}_c (1/2^-, 1^-)$, there is one form factor.
For the decay to ${\tilde \Omega}_c (1/2^-, 0^-)$,
there is also one form factor. Using the
method given in \cite{r3,r4}, one can write down
form factor structures in these decays.
For $\Omega_b \rightarrow {\tilde \Omega}_c (1/2^-, 0^-)$
\FL
\begin{eqnarray}
\langle {\tilde \Omega}_c (1/2^-, 0^-) |\bar b\ \Gamma\ c |
\Omega_b \rangle
={\bar u}_c (v^\prime)\ \Gamma B^\nu_b(v) \
\xi_3\ (v-v^\prime)_\nu \
\label{eq:e7}
\end{eqnarray}
where $u_c$ indicates the spinor of
${\tilde \Omega}_c (1/2^-, 0^-)$ and
$B^\nu_b(v)$ is the same as defined in Eq.~(\ref{eq:e5}).
For $\Omega_b \rightarrow {\tilde \Omega}_c (3/2^-, 1^-)$ or
${\tilde \Omega}_c (1/2^-, 1^-)$,
\FL
\begin{eqnarray}
\langle {\tilde \Omega}_c (3/2^-, 1^-)\ {\rm or}
&&{\tilde \Omega}_c (1/2^-, 1^-) |\bar b\ \Gamma\ c |
\Omega_b \rangle=\nonumber\\
&&{\bar B}^\mu_c (v^\prime)\ \Gamma\ B^\nu_b(v)
\ \xi_4\ \epsilon_{\mu\nu\alpha\beta} v^\alpha v^{\prime\beta}
\label{eq:e8}
\end{eqnarray}
where $B^\mu_c(v^\prime)$ is defined for
${\tilde \Omega}_c (3/2^-, 1^-)$ or
${\tilde \Omega}_c (1/2^-, 1^-)$ similarly as in Eq.(\ref{eq:e5}).
For $\Omega_b \rightarrow {\tilde \Omega}_c (5/2^-, 2^-)$ or
${\tilde \Omega}_c (3/2^-, 2^-)$,
\FL
\begin{eqnarray}
&&\langle {\tilde \Omega}_c (5/2^-, 2^-)
\ {\rm or}\ {\tilde \Omega}_c (3/2^-, 2^-) |\bar b\ \Gamma\ c |
\Omega_b \rangle=\nonumber\\
&&{\bar B}^{\mu\lambda}_c (v^\prime) \Gamma B^\nu_b(v) \
v^\lambda (\xi_5\ g_{\mu\nu}+\xi_6\ (v-v^\prime)_\mu
(v-v^\prime)_\nu ) \
\label{eq:e9}
\end{eqnarray}
For ${\tilde \Omega}_c (5/2^-, 2^-)$,
$B^{\mu\lambda}_c(v^\prime)$ is just the spin-5/2
Rarita-Schwinger spinor-vector $u^{\mu\lambda}_c$. For
${\tilde \Omega}_c (3/2^-, 2^-)$ $B^{\mu\lambda}_c(v^\prime)$ is
defined as
\FL
\begin{eqnarray}
B^{\mu\lambda}_c (v^\prime)\!\!=\!\!
\frac{ (\gamma^\mu+v^{\prime\mu})\gamma_5
u^\lambda_c(v^\prime) }{\sqrt{10}}+
\frac{ (\gamma^\lambda+
 v^{\prime\lambda})\gamma_5 u^\mu_c(v^\prime) }{\sqrt{10}}
\label{eq:e10}
\end{eqnarray}
where $u^\mu_c$ is the spin-3/2 Rarita-Schwinger spinor-vector.

In the above discussion we have confined ourselves to excited
states with only angular excitations. One can easily include
excited states with radial excitations.
These states can be generally denoted by
${\tilde \Omega}^{(n)}_c (S^\pi, s_l^{\pi_l})$. Thus we
consider generally
${\tilde \Omega}^{(n)}_c (5/2^-, 2^-)$,
${\tilde \Omega}^{(n)}_c (3/2^-, 2^-)$,
${\tilde \Omega}^{(n)}_c (3/2^-, 1^-)$,
${\tilde \Omega}^{(n)}_c (1/2^-, 1^-)$ and
${\tilde \Omega}^{(n)}_c (1/2^-, 0^-)$. The form factor
structures involving these states are the same as those given in
Eq.~(\ref{eq:e7}) to Eq.~(\ref{eq:e10}) except that one now
needs to add a index $n$ to the form factors to indicate the radial
excitation, i.e.
$\xi_i \rightarrow \xi_i^{(n)}$ ($i=3,4,5,6$). Similarly,
the form factor structures involving
excited states with no angular excitation but radial excitation
are the same as those in Eq.~(\ref{eq:e4}) except one has to
make the substitution $\xi_i\rightarrow (w-1)\xi_i^{(n)}$
($i=1,2$).

Having established the Lorentz structures, it is then
straightforward to derive the Bjorken sum rule
for $\Omega_b$ semiletponic decays using a
formula similar to Eq.~(\ref{eq:e1}).
In so doing one needs to use the projection operators.
For spin 1/2 baryons,
\FL
\begin{eqnarray}
\sum_{s} u(v,s) \bar u(v,s)=(1+\not\!v)
\label{eq:e11}
\end{eqnarray}
For spin 3/2 baryons,
\FL
\begin{eqnarray}
&&\sum_{s} u^\mu(v,s) \bar u^\nu(v,s)=(1+\not\!v)\nonumber\\
&&\left\{ -g^{\mu\nu}+\frac{2}{3}\ v^\mu v^\nu+
\frac{1}{3}\ \gamma^\mu \gamma^\nu+
\frac{1}{3} (\gamma^\mu v^\nu-\gamma^\nu v^\mu ) \right\}
\label{eq:e12}
\end{eqnarray}
For spin 5/2 baryons \cite{r8a},
\widetext
\FL
\begin{eqnarray}
&&\sum_{s} u^{\alpha_1\alpha_2}(v,s)
\bar u^{\alpha_3\alpha_4}(v,s)
=(1+\not\!v) \biggr\{ \
\frac{2}{5} v^{\alpha_1} v^{\alpha_2}
v^{\alpha_3} v^{\alpha_4}
\nonumber\\
&&+\frac{1}{5}
[\  v^{\alpha_3} v^{\alpha_4}
(v^{\alpha_1} \gamma^{\alpha_2}+v^{\alpha_2} \gamma^{\alpha_1})
- v^{\alpha_1} v^{\alpha_2}
(v^{\alpha_4} \gamma^{\alpha_3}+v^{\alpha_3} \gamma^{\alpha_4}) \ ]
\nonumber\\
&&+\frac{1}{5}
[\  v^{\alpha_3} v^{\alpha_4} g^{\alpha_1\alpha_2}+
  v^{\alpha_1} v^{\alpha_2} g^{\alpha_3\alpha_4} \ ]
-\frac{2}{5}
[\  v^{\alpha_3} v^{\alpha_1} g^{\alpha_2\alpha_4}+
  v^{\alpha_4} v^{\alpha_1} g^{\alpha_2\alpha_3}+
  v^{\alpha_3} v^{\alpha_2} g^{\alpha_1\alpha_4}+
  v^{\alpha_4} v^{\alpha_2} g^{\alpha_1\alpha_3} \ ]
\nonumber\\
&&+\frac{1}{10}
[\  v^{\alpha_1} v^{\alpha_3} \gamma^{\alpha_2} \gamma^{\alpha_4}+
  v^{\alpha_1} v^{\alpha_4} \gamma^{\alpha_2} \gamma^{\alpha_3}+
  v^{\alpha_2} v^{\alpha_3} \gamma^{\alpha_1} \gamma^{\alpha_4}+
  v^{\alpha_2} v^{\alpha_4} \gamma^{\alpha_1} \gamma^{\alpha_3} \ ]
\nonumber\\
&&-\frac{1}{10}
[\  g^{\alpha_1\alpha_3} \gamma^{\alpha_2} v^{\alpha_4}+
  g^{\alpha_1\alpha_4} \gamma^{\alpha_2} v^{\alpha_3}+
  g^{\alpha_2\alpha_3} \gamma^{\alpha_1} v^{\alpha_4}+
  g^{\alpha_2\alpha_4} \gamma^{\alpha_1} v^{\alpha_3} \ ]
\nonumber\\
&&+\frac{1}{10}
[\  g^{\alpha_1\alpha_3} v^{\alpha_2} \gamma^{\alpha_4}+
  g^{\alpha_1\alpha_4} v^{\alpha_2} \gamma^{\alpha_3}+
  g^{\alpha_2\alpha_3} v^{\alpha_1} \gamma^{\alpha_4}+
  g^{\alpha_2\alpha_4} v^{\alpha_1} \gamma^{\alpha_3} \ ]
\nonumber\\
&&-\frac{1}{10}
[\  g^{\alpha_1\alpha_3} \gamma^{\alpha_2} \gamma^{\alpha_4}+
  g^{\alpha_1\alpha_4} \gamma^{\alpha_2} \gamma^{\alpha_3}+
  g^{\alpha_2\alpha_3} \gamma^{\alpha_1} \gamma^{\alpha_4}+
  g^{\alpha_2\alpha_4} \gamma^{\alpha_1} \gamma^{\alpha_3} \ ]
\nonumber\\
&&-\frac{1}{5} g^{\alpha_1\alpha_2} g^{\alpha_3\alpha_4}
+\frac{1}{2}[\  g^{\alpha_1\alpha_3} g^{\alpha_2\alpha_4}+
                g^{\alpha_1\alpha_4} g^{\alpha_2\alpha_3} \ ]
\ \biggr\}
\label{eq:e13b}
\end{eqnarray}
\narrowtext
\noindent We finally obtain the Bjorken sum rule
\widetext
\FL
\begin{eqnarray}
&&1=\frac{(2+w^2)}{3} |\xi_1|^2+\frac{ (w^2-1)^2}{3} |\xi_2|^2
+\frac{(w-w^3)}{3}(\xi_1 \xi_2^\ast+\xi_2\xi_1^\ast)\nonumber\\
&&+(w-1) \sum_{n} \biggr\{\
\frac{(w+1)}{3} |\xi_3^{(n)}|^2+\frac{10(w+1)}{9} |\xi_4^{(n)}|^2
+[\ \frac{ (w+1)(2w^2+3)}{9} |\xi_5^{(n)}|^2\nonumber\\
&&+\frac{2(w+1)(w^2-1)^2}{9}|\xi_6^{(n)}|^2
-\frac{2w(w+1)(w^2-1)}{9}(
\xi_5^{(n)} \xi_6^{(n)\ast}+\xi_6^{(n)} \xi_5^{(n)\ast} )\ ]
\nonumber\\
&&+(w-1) [ \ \frac{(2+w^2)}{3} |\xi_1^{(n)}|^2+
\frac{ (w^2-1)^2}{3} |\xi_2^{(n)}|^2
+\frac{(w-w^3)}{3} (\xi_1^{(n)} \xi_2^{(n)^\ast}+
\xi_2^{(n)} \xi_1^{(n)^\ast})\ ] \biggr\}\nonumber\\
&&+...
\label{eq:e14}
\end{eqnarray}
\narrowtext
\noindent where the ellipsis denotes the contributions which have
not been included here. These contributions are always positive.
Obviously, this sum rule is consistent with Eq.~(\ref{eq:e6}). We
expect Eq.~(\ref{eq:e14}) to be valid near $w=1$. Note that
time reversal invariance will require, for example,
$\xi_1$ and $\xi_2$  to be relatively real.

The Bjorken sum rule from Eq.~(\ref{eq:e14})
gives an upper limit to the form factors.
For example,
\FL
\begin{eqnarray}
&&\!1\!\geq \!\!\frac{2+w^2}{3} |\xi_1|^2+
\!\!\frac{ (w^2-1)^2}{3} |\xi_2|^2
+\!\!\frac{(w-w^3)}{3} (\xi_1\xi_2^\ast+\xi_2 \xi_1^\ast)
\nonumber\\
&& \
\label{eq:e15}
\end{eqnarray}
The sum rule in Eq.~(\ref{eq:e14}) also provides a restriction
on the slope of the form factor. Expanding $\xi_1$ and $\xi_2$
near $w=1$
\FL
\begin{eqnarray}
\xi_i=\xi_i(1)-(w-1) \theta_i+...  \ \ \ (i=1,2)
\label{eq:e18}
\end{eqnarray}
one can derive from Eq.~(\ref{eq:e14})
\FL
\begin{eqnarray}
\theta_1 =&& \frac{1}{3}-\frac{2}{3} \xi_2(1)\nonumber\\
\!\!+&&\!\!\sum_{n}(\ \frac{1}{3}|\xi_3^{(n)}(1)|^2+
\frac{10}{9}|\xi_4^{(n)}(1)|^2
+\frac{5}{9}|\xi_5^{(n)}(1)|^2 \  )\nonumber\\
+&&...
\label{eq:e19}
\end{eqnarray}
Since the contributions from all
excited states are positive,
\begin{eqnarray}
\theta_1 \geq \frac{1}{3}-\frac{2}{3} \xi_2(1)
\label{eq:e20}
\end{eqnarray}

It is obvious that all the above results are also valid
for the corresponding $\Sigma_b$ semileptonic decays.

The restriction on the form factors involving the ground states,
Eq.~(\ref{eq:e15}), is generally not very strong restriction.
However,
one can compare it with the form factors calculated or assumed in
some models \cite{r8,r9,r10} on
$\Omega_b \rightarrow (\Omega_c, \Omega_c^\ast)$ semileptonic
decays. In \cite{r8} and \cite{r9}, the form factors are
parametrized by $f$ and $g$ which are related to $\xi_1$ and
$\xi_2$ through: $\xi_1=\frac{\textstyle w+1}{\textstyle2} f+g$,
$\xi_2=\frac{\textstyle f}{\textstyle 2}$.
$g=0$ is set in these models. Thus Eq.~(\ref{eq:e15}) implies
that $1\geq \frac{\textstyle (1+w)^2}{\textstyle 4} f^2$,
i.e. $f$ should be smaller than given by a
monopole $q^2$-dependence. Note a monopole $q^2$-dependence
normalized to unity at $w=1$ has just the form
$\frac{\textstyle 2}{\textstyle 1+w}$
in the heavy quark limit. The form factor $f$ used in \cite{r8}
and \cite{r9} are consistent with this restriction.

In \cite{r10}, the form factors in
$\Sigma_b \rightarrow \Sigma_c e \bar \nu_e$ are
calculated in the so-called quark confinement model. In the
heavy quark limit the restriction in eq.~(\ref{eq:e15}) is
also valid for these form factors.
$\xi_1$ and $\xi_2$ in this model are given by \cite{r8}
\begin{eqnarray}
&&\xi_1=w \Psi \ \ \ \ \xi_2=\Psi
\label{eq:e16a}\\
&&\Psi=\frac{ ln(\omega+\sqrt{\omega^2-1})}{\sqrt{\omega^2-1}}
\label{eq:e16b}
\end{eqnarray}
Substituting  Eq.~(\ref{eq:e16b}) into~(\ref{eq:e15}), one gets
\begin{eqnarray}
\frac{3}{1+2 \omega^2} \geq \Psi^2
\label{eq:e17}
\end{eqnarray}
It is easy to show that
$\Psi$ of Eq.~(\ref{eq:e16b}) given in \cite{r10}
disobeys the upper limit for $\Psi$.

In summary, we have derived the Bjorken sum rule for
$\Omega_b$ (or $\Sigma_b$) semileptonic decays to ground and
low-lying negative-partiy excited charmed states,
in the heavy quark limit.
The sum rule gives restriction on the form factors.
We have compared this restriction
with some models for $\Omega_b$ semileptonic decays and found that
not
all models obey this restriction.

\acknowledgments

This research was supported by a grant from the
Natural Sciences and Engineering Research Council of Canada to
A. N. Kamal.
The author is very grateful to A. N. Kamal for stimulating
discussions and suggestions.

\end{document}